\documentclass[12pt,aps,prb,preprint]{revtex4}   % style for Physical Review B and AJP are similar

\usepackage{amsmath}    % need for subequations
\usepackage{amsfonts}  %note how statements can be commented out
\usepackage{amssymb}
\usepackage{graphicx}   % for figures
\begin{document}

\title{Tipping of a classical point mass pendulum: Role of statistical fluctuations}
\author{Abhishodh Prakash}
\email{abhishodh@gmail.com}   %optional 
\altaffiliation[Also at ]{Birla Institute of Technology and Science- Pilani, KK Birla Goa Campus, Zuarinagar, India 403726}  %  optional
\affiliation{Raman Research Institute, Bangalore, India 560080}

\date{\today}

\begin{abstract}
The behavior of a stationary inverted point mass pendulum pivoted at its lower end in a gravitational potential is studied under the influence of statistical fluctuations. It is shown using purely classical equations that the pendulum eventually tips over i.e evolves out of its initial position of unstable equilibrium, and, in a finite amount of time points down assuming a position of stable equilibrium. This `tipping time' is calculated by solving the appropriate Fokker- Planck equation in the overdamped limit. It is also shown that the asymptotic time solution for probability corresponds to the Boltzmann distribution, as expected for a system in stable equilibrium, and that the tipping time tends to infinity as the parameter corresponding to the strength of thermal fluctuations is tuned to zero, thereby defining the limit where one recovers the classical result that a stationary inverted point mass pendulum never tips over. The paper provides a unique perspective showing that phenomena like tipping that have been often attributed to quantum mechanics can be studied even in the domain of purely classical physics.
\end{abstract}

\maketitle

\section{Introduction}
Inverted stationary pendulums have always been an interesting study. Classically, in the absence of any perturbations to the gravitational potential, such a position of unstable equilibrium corresponds to a saddle node in phase space \cite{strogatz}. If the system is conservative i.e there is no damping, these saddle nodes are connected by heteroclinic trajectories of infinite time period. In the presence of damping, these trajectories change to spirals and fall into stable fixed points. In both cases, however, as long as the system is left undisturbed, it stays precariously at the saddle node. Simply put, from such a classical standpoint, an unperturbed stationary inverted point mass pendulum stays inverted forever. The scenario changes when one seeks a quantum mechanical solution. The uncertainty principle forbids fixing both the position and velocity simultaneously. One can at best localize the initial position of the pendulum within a `small enough' region around the inverted position and study its behavior. Work has been done in the past\cite{op,pend,rod} to explore how the uncertainty principle causes quantum rods and pendulums to get delocalized and tip over.

So far, this tipping has been well studied as a quantum mechanical phenomenon. Considering a pendulum with friction, say in a jar of olive oil, it is hard to believe that tipping can happen only through quantum effects. Even in the classical world, the fluctuation- dissipation theorem tells us that in a dissipative system, there are fluctuations of a statistical rather than quantum origin which can very well push the pendulum out of equilibrium. In this paper aimed at undergraduate students, graduate students and teachers of physics, we return to the classical pendulum, study its evolution in the presence of finite temperature statistical fluctuations and show that even classical effects can lead to such interesting behavior. The pendulum is classical in the sense that its initial position can be exactly fixed to the inverted position of unstable equilibrium with zero momentum. We work in the high friction (overdamped) limit and use the corresponding Fokker- Planck equation for this system to study its behavior and quantify the time it takes to tip over. 

\section{Formulation of the problem}

Consider a point body of mass `m' fixed to one end of a massless rigid rod of length `l', while the other end is pivoted. Let this system be free to move in a vertical plane under the influence of a gravitational potential of the form:
\begin{equation} \label{potential}
V(\theta) = mgl\cos\theta, ~\theta \in [0,2 \pi]
\end{equation}
where $\theta$ is measured from the vertical. Now, we introduce white noise from statistical fluctuations in the form of a stochastic variable, $\xi(t)$. The noise is white in the sense that it is delta correlated in time:
\begin{equation}\label{deltacorrelation}
\langle \xi(t) \xi(t') \rangle = 2\gamma k_B T~ \delta(t-t')
\end{equation}
where, $\gamma$ is the coefficient of friction, $k_B$ is the Boltzmann constant and $T$ is the absolute temperature\cite{reichl}. With these, we can write the `Langevin Equations' of motion for the pendulum as:
\begin{subequations} \label{lang1}
\begin{eqnarray}
(ml^2) \dfrac{d \dot{\theta}(t)}{dt} = -\gamma \dot{\theta}(t) -\dfrac{dV(\theta)}{d\theta} + \xi(t)\\
\dot{\theta}(t) = \dfrac{d\theta(t)}{dt} 
\end{eqnarray}
\end{subequations}
We work in the high friction limit and linearize the above equations by dropping the intertia term to get,
\begin{equation} \label{lang}
\dfrac{d\theta(t)}{dt} = \dfrac{1}{\gamma} \left(-\dfrac{dV(\theta)}{d\theta} + \xi(t)\right)
\end{equation}
Using Eqs.~\ref{lang}, \ref{potential} and \ref{deltacorrelation}, we can expect to obtain a formal solution to the expectation value $\langle \theta(t) \rangle$. 

Equivalently, we can work with probabilities. Defining $\rho(\theta,t)$ as the probability density of finding the pendulum at position `$\theta$' at time `$t$' and using Eqs.~\ref{lang} and \ref{deltacorrelation} in the continuity equation associated with $\rho(\theta,t)$, we can find the evolution of the expectation value $\langle \rho(\theta,t) \rangle =P(\theta,t)$. The `Fokker- Planck' equation (corresponding to Eq.~\ref{lang}) that describes this evolution can be written as:
\begin{equation} \label{fokker}
\dfrac{\partial P(\theta,t)}{\partial t} = \dfrac{1}{\gamma} \dfrac{\partial}{\partial \theta} \left( \dfrac{dV(\theta)}{d\theta} P(\theta,t) + k_BT \dfrac{\partial P(\theta,t)}{\partial \theta} \right) 
\end{equation}
The steps involved in moving from Eq.~\ref{lang} to Eq.~\ref{fokker} can be found in Ref~.\onlinecite{reichl}. We prefer to work in the probability picture and perform our analysis with Eq.~\ref{fokker}.

\section{Numerical solution to the Fokker- Planck equation and estimation of the tipping time}

Substituting Eq.~\ref{potential} and defining the dimensionless quantities $\tau = \dfrac{mgl}{\gamma} t$ and $\alpha=\dfrac{k_BT}{mgl}$ we can rewrite Eq.~\ref{fokker} as:
\begin{equation}\label{fokkerproper}
\dfrac{\partial P(\theta,\tau)}{\partial \tau} = -P(\theta,\tau) \cos\theta- \dfrac{\partial P(\theta,\tau)}{\partial \theta} \sin\theta + \alpha~\dfrac{\partial^2P(\theta,\tau)}{\partial \theta^2}
\end{equation}
Henceforth, we shall work with the dimensionless time `$\tau$' and any reference to `time' would mean the latter. Also, `$\alpha$' can be regarded as the strength of thermal fluctuations. We assume that at $\tau = 0$, the pendulum is exactly at $\theta=0$ i.e:
\begin{equation}\label{bc1}
P(\theta,0)=\delta(\theta)
\end{equation}
Also, since the pendulum moves in a circular space, we impose periodicity in $\theta$:
\begin{equation}\label{bc2}
P(\theta,\tau)=P(\theta+2\pi,\tau)
\end{equation}
Using Eq.~\ref{fokkerproper}, with the boundary conditions Eq.~\ref{bc1} and Eq.~\ref{bc2}, we can obtain a solution for $P(\theta,\tau)$. 

To do this analytically we would have to obtain the solution as an eigenfunction expansion by spectral resolution of the differential operator on the right hand side of Eq.~\ref{fokkerproper}. First however, the operator has to be made self adjoint by using a suitable transformation on the independent variable. A detailed prescription to do this can be found in Ref.~\onlinecite{reichl}. Such a calculation is mathematically cumbersome. We prefer to simply solve the problem numerically and find that it captures the essence of the solution effectively.

\subsection{Asymptotic time solution}
We can calculate the solution ($P(\theta,\infty) = \Phi(\theta)$) for Eq.~\ref{fokkerproper} which is the probability distribution when the system has completely relaxed to its stable configuration, by setting the time derivative to zero:
\begin{equation}\label{asymp}
-\Phi(\theta) \cos\theta- \dfrac{d\Phi(\theta)}{d\theta} \sin\theta + \alpha~\dfrac{d^2\Phi(\theta)}{d\theta^2} = 0
\end{equation}
As one expects for a system in stable equilibrium, the solution for Eq.~\ref{asymp} considering Eq.~\ref{bc1} and Eq.~\ref{bc2} is a normalized Boltzmann distribution centred around $\theta = \pi$.
\begin{equation}\label{boltzman}
\Phi(\theta) = \left(2 \pi  I_0\left(\frac{1}{\alpha}\right)\right)^{-1}~ exp\left(-\frac{\cos\theta}{\alpha}\right) 
\end{equation}
where $I_0$ is the zero order modified Bessel function of the first kind and the prefactor comes from normalizating $\Phi(\theta)$ on the circle. We expect that long after tipping, the probability distribution assumes the form of $\Phi(\theta)$ displayed in Eq.~\ref{boltzman}.

\subsection{Evolution of probability density}

Fig.~\ref{distroevolution} shows the evolution of the spatial probability distribution with time. It can be seen that with time, probability oozes out of the initial delta peaks, the maximal bumps travel towards $\theta = \pi$ and upon reaching it, stay there. It is also seen that the asymptotic probability density indeed corresponds to $\Phi(\theta)$ of Eq.~\ref{boltzman}. Another noteworthy observation is the symmetry of the solutions about $\theta = \pi$. This makes physical sense as the pendulum is equally likely to tip clockwise as it is to tip anti- clockwise and during the actual process of tipping, the pendulum breaks this symmetry. Mathematically, the fact that the probability solutions respect parity can be confirmed by observing that Eq.~\ref{fokkerproper} is invariant under the transformation $\theta \rightarrow - \theta$. 

\begin{figure}[h]
\begin{center}
\includegraphics{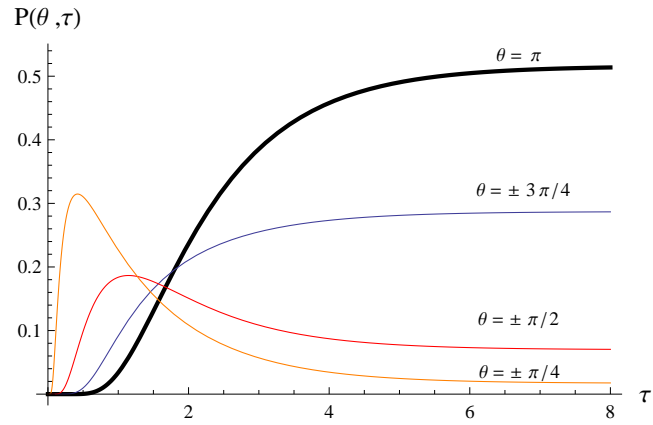}
\caption{\protect\label{timeevolution}Evolution of $P(\theta,\tau)$ at various $\theta$ setting $\alpha=0.5$ }
\end{center}
\end{figure}

Fig.~\ref{timeevolution} shows the time evolution of $P(\theta,\tau)$ at various angles. It is seen that the graphs do not oscillate which is typical of solutions in the high friction limit when the inertia term is neglected. Once again, we see that at large times, the probability density is maximum for $\theta = \pi$. This confirms that after a very long time, the pendulum is most likely to be found at the stable equilibrium point of $\theta = \pi$. One can thus define the tipping time, $\tau_{tip}$ as the time when the probability is first a global maximum at $\theta = \pi$. 

\subsection{Calculating the tipping time and recovering the classical mechanical limit}
\begin{figure}[h]
\begin{center}$
\begin{array}{cc}
\includegraphics[width=2.5in]{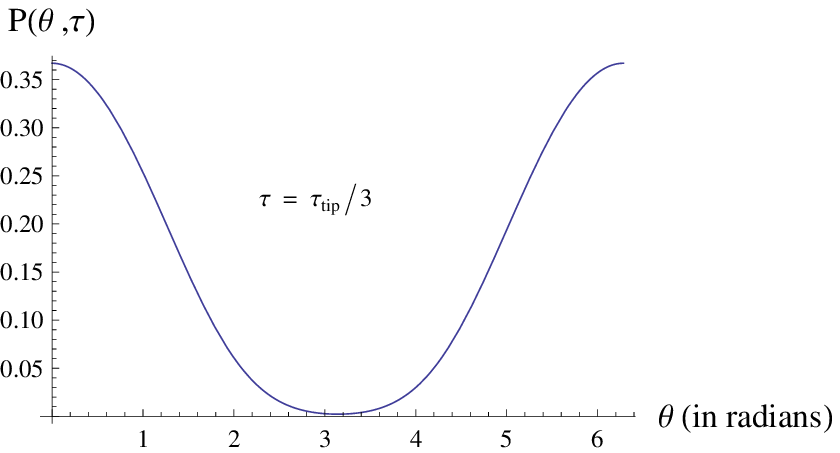} &
\includegraphics[width=2.5in]{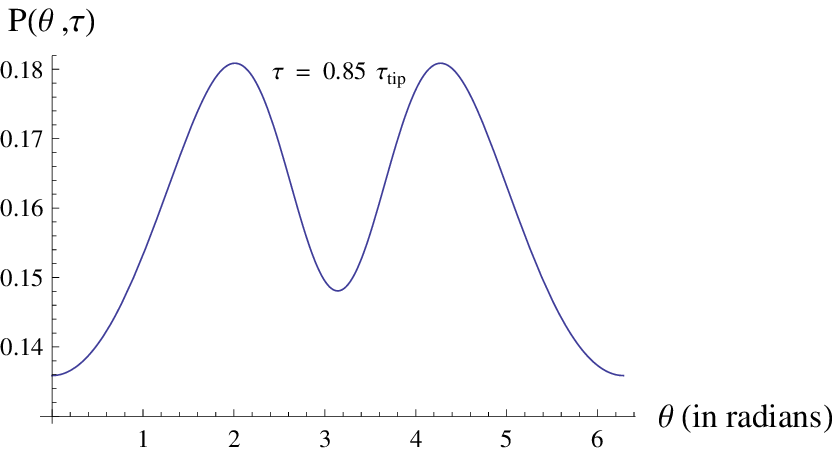} \\ 
\includegraphics[width=2.5in]{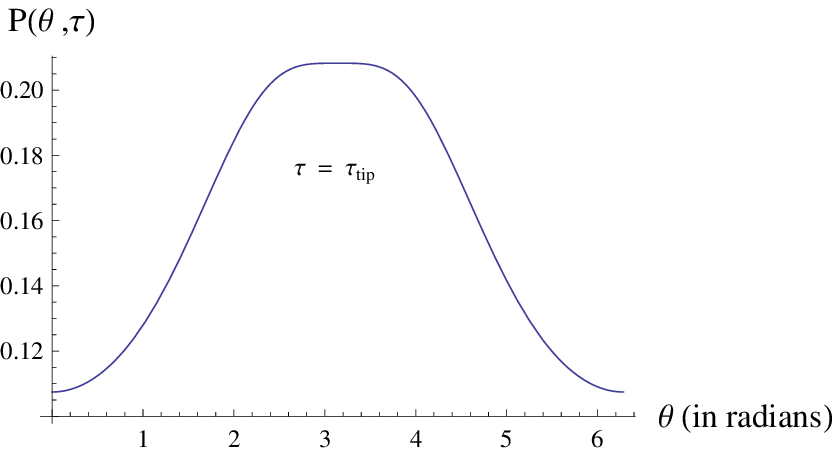} &
\includegraphics[width=2.5in]{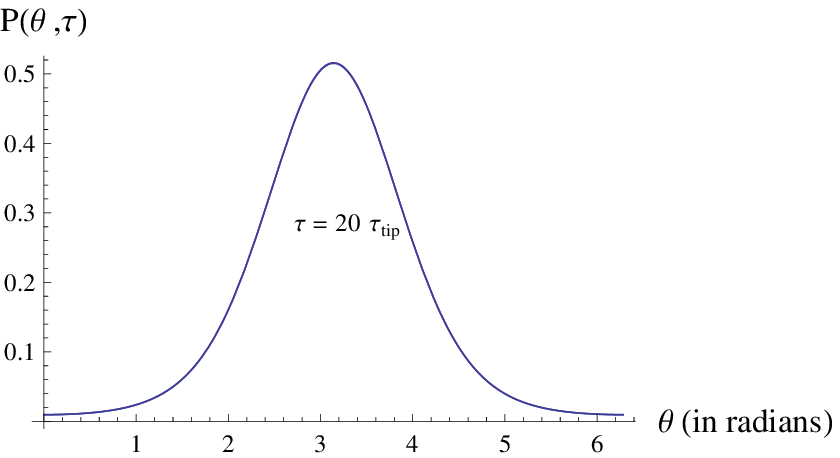}
\end{array}$
\end{center}
\caption{\protect\label{distroevolution}Evolution of $P(\theta,\tau)$ versus $\theta$ distribution setting $\alpha=0.5$}
\end{figure}

\begin{figure}[h]
\begin{center}
\includegraphics{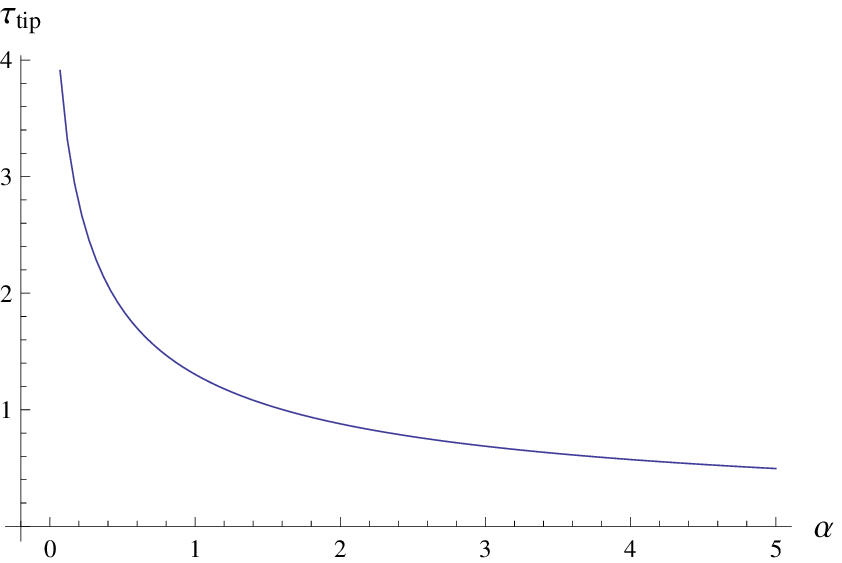}
\caption{\protect\label{tippingtime}Tipping time, $\tau_{tip}$ versus thermal strength, $\alpha$ }
\end{center}
\end{figure}

Defining the tipping time as the instant when probability is first maximum at $\theta = \pi$ suggests that the value of $\tau_{tip}$ can be calculated by solving the equation
\begin{equation}\label{redundanttip}
\dfrac{\partial P(\pi,\tau_{tip})}{\partial \theta} = 0
\end{equation}
However, observing Fig.~\ref{distroevolution} we see that $\theta=\pi$ is always a point of local extremum and therefore such a slope is always zero. This arises from the symmetry about $\theta=\pi$ mentioned earlier. On the other hand, we can see from Fig.~\ref{distroevolution} that as $\theta=\pi$ makes a transition from being a point a local minimum to global maximum, the second derivative changes sign from positive to negative. Thus, we can recognize $\tau_{tip}$ as the instant when the second derivative becomes zero, by solving the equation
\begin{equation}\label{redundanttip}
\dfrac{\partial^2 P(\pi,\tau_{tip})}{\partial \theta^2} = 0
\end{equation}
Fig.~\ref{tippingtime} shows the variation of $\tau_{tip}$ with $\alpha$. It can be seen that when $\alpha\rightarrow \infty$ i.e when the thermal fluctuations greatly dominate, the pendulum tips instantaneously ($\tau_{tip}\rightarrow0$). On the other hand, when $\alpha\rightarrow 0$, we find that $\tau_{tip}\rightarrow \infty$ which is tantamount to saying that in the limit of negligible thermal fluctuations, we recover the limit in which the pendulum never tips over.  

Finally, let us calculate the tipping time for a typical system to get a feel for the order of numbers involved. Considering a point mass pendulum of length $l = 10 \mu m$ with a spherical pollen grain of mass $m = 10^{-14}kg$ and radius $r =1 \mu m $ acting as the point mass, immersed in olive oil with coeffecient of dynamic viscocity $\eta = 0.1 Pa.s$ at room temperature ($T = 300 K$), we can calculate $\tau_{tip}$ for this configuration which corresponds to $\alpha = 0.0042$ as 3.206 dimensionless units. Comparing the definition of $\gamma$ in Eq.~\ref{lang} and the formal definition\cite{hcverma} of $\eta$, it is easy to see that the two are linked by $\gamma = \eta(\pi r^2) l$. With this, we can calculate the observable tipping time $t_{tip} = \dfrac{\gamma}{mgl} \tau_{tip}$ as 21.919 seconds. 

\section{Summary and conclusion}
The evolution of an initially stationary and inverted classical pendulum in the presence of statistical fluctuations is studied and it is found that these fluctuations push the pendulum out of unstable equilibrium. The time taken for the pendulum to tip to the configuration of stable equilibrium is calculated numerically and it is seen that within the confines of classical analysis, only in the limit of negligible thermal fluctuations does the pendulum stay inverted forever. Analysis of this simple system elegantly demonstrates physical phenomena that commonly occur in dynamical systems out of stable equilibrium and hence can be used not only in pedagogy but also as an effective toy model \cite{mazenko} to study complicated systems and their behavior, in wider and more profound contexts. Also, the analysis being purely classical demonstrates the abundance of highly interesting physics in the relatively forgotten realm of classical physics that are still unexplored and are worthwhile to study to gain a rich insight into the physical laws governing observable reality.

\begin{acknowledgments}
I thank Joseph Samuel and Supurna Sinha for their valuable guidance, discussions and suggestions on improving the manuscript. I also thank Onkar Parrikar for the various discussions that eventually developed into the idea behind this paper. 
\end{acknowledgments}

\newpage
\section*{Figures}

\begin{thebibliography}{5}

\bibitem{strogatz} Steven H. Strogatz, \textsl{Nonlinear Dynamics and Chaos} (Levant, Kolkata, India, 2007), Indian ed,Chap. 6.

\bibitem{op}Onkar Parrikar,
``Tipping time of a quantum rod'' Eur. J. Phys. {\bf 31}, 317--324 (2010).

\bibitem{pend}G.P.Cook and Clyde S. Zaidins,
``The quantum point mass pendulum'' Am. J. Phys. {\bf Vol. 54, No. 3}, 259--261 (March 1986).

\bibitem{rod}M. Batista and J. Peternelj,
``Quantum cards and quantum rods'' arXiv:quant-ph/0611036v1 (2006)

\bibitem{reichl} L.E.Reichl, \textsl{A Modern Course in Statistical Physics} (Wiley-Interscience, Kolkata, India, 2007), 2nd. ed.

\bibitem{mazenko} For instance, the symmetry breaking aspect was explored by Gene F. Mazenko, William G. Unruh and Robert M. Wald in
``Does a phase transition in the early universe produce the conditions needed for inflation'' Phys. Rev. D. {\bf Vol. 31, No. 2}, 273--282 (January 1985).

\bibitem{hcverma}Can be found in most text books on introductory physics like H.C.Verma, \textsl{Concepts of Physics, Vol 1} (Bharati Bhavan Publishers and Distributors, Patna, India, 2003), 2nd reprint.

\end{thebibliography}
\end{document}